\documentclass[aps,pre,twocolumn,showpacs,preprintnumbers,amsmath,amssymb]{revtex4}
\usepackage{epsfig}

\begin{document}
\title{External bias in the model of isolation of communities}
\author{Julian Sienkiewicz$^1$, Grzegorz Siudem$^{1,2}$, and Janusz A. Ho{\l}yst$^1$}
\affiliation{$^1$Faculty of Physics, Center of Excellence for Complex Systems Research, Warsaw University of Technology, Koszykowa 75, PL-00-662 Warsaw, Poland\\
$^2$Faculty of Mathematics and Information Science, Warsaw University of Technology, Pl. Politechniki 1, PL-00-661 Warsaw, Poland}
\date{\today}

\begin{abstract}
We extend a model of community isolation in the $d$-dimensional lattice onto the case with an imposed imbalance between birth rates of competing communities. We give analytical and numerical evidences that in the asymmetric two-specie model there exists a well defined value of the asymmetry parameter when the emergence of the isolated (blocked) subgroups is the fastest, i.e. the  characteristic time $t_c$ is minimal. This critical value of the parameter depends only on  the lattice dimensionality and is independent from the system size. Similar phenomenon was observed in the multi-specie case with a geometric distribution of the birth rates. We also show that blocked subgroups in the multi-specie case are absent or very rare when either there is a strictly dominant specie that outnumbers the others or when there is a large diversity of species. The number of blocked species of different kinds decreases with the dimension of the multi-specie system.
\end{abstract} 
\pacs{05.50.+q, 89.75.Hc, 02.50.-r} \maketitle

\section{INTRODUCTION}
The question of {\it imbalance} or {\it asymmetry} is long known in many popular and fundamental phenomena in non-equilibrium statistical physics such as gradient-induced transfer of particles, heat or current \cite{statphys_book}. The two-specie fermion mixtures and the impact of its population imbalance on the phase diagram of uniform superfluidity is also one of the multiple faces of the mentioned problem \cite{iskin_prl,dao_prl}.

This issue emerges in a quite natural way in other fields: for example in ecology, where the imbalance can be related to inclusion of the non-native species in certain area \cite{pranovi_binv}, as well as to the absolute number of species in the branches of phylogenetic tress \cite{holman} or in general it can govern the whole evolution in a specific ecosystem \cite{lotka_volterra}. In economics the heterogeneity can be manifested for example by Zipf's law in distribution of firm sizes \cite{stanley_nature,axtell_science} or in the statistics of order books \cite{zaccaria_pre}. 

Apart from the above mentioned quantitative sciences, there is also sociology, during the last two decades under frequent consideration of physicists which eventually led to the emergence of novel scientific entity---sociophysics \cite{castellano_rmp}. Its powerful tools have been widely used not only to face the problems of culture spreading \cite{axlerod}, collective behavior of audience \cite{farkas_nature}, correspondence activity \cite{vazquez_prl} or travel customs \cite{brockmann_nature,gonzalez_nature} but very recently it also touches the issues of moral standards \cite{helbing_plos} and emotions \cite{sobkowicz_epjb,czaplicka_app,chmiel_ijmpc}. By the same token, a significant attention is devoted to {\it social balance}. Its evolution on networks with both friendly and unfriendly relations has proven to provide signs of phase transitions \cite{antal_physad} or complex energy landscapes \cite{marvel_prl}. The complexity is inseparably binded with human nature with respect to the {\it social imbalance}: it is true that people are inequality-aware which is backed even by neural evidence \cite{tricomi_nature} but then again they also tend to accept inequalities reflecting differences in individual achievements \cite{almas_science}.

In this paper we draw attention to the issue of imbalance in a model of isolation of species (communities) extending the previously obtained results \cite{sienkiewicz_pre}. Social isolation is a crucial problem which can be caused by various factors such as illness, imprisonment or emigration and is regarded to have a substantial impact on higher suicide rates \cite{roehner_book}. In this context, isolation may also emerge as an effect of the social imbalance, just to mention large discrepancies between number of males and females in France after World War I \cite{roehner_book}. Bearing that in mind, we shall try to show the impact that different methods of imbalance introduction may have on the number of isolated species.

The paper's structure is organized as follows: in Sec. \ref{sec:basic_model} we describe in short the basic model of isolation, Sec. \ref{sec:gen_desc} gives details of the ways the imbalance is introduced and Sec. \ref{sec:num} presents exhaustive analysis of the numerical simulations and the theoretical approach. The outcome is summarized in Sec. \ref{sec:conc} along with some general remarks.
\section{BASIC MODEL}\label{sec:basic_model}
Recently we introduced a simple model of community isolation \cite{sienkiewicz_pre} whose basic rules can be described in the following way: in each time step one puts a representative of  specie ($\uparrow$) or ($\downarrow$) in a random, unoccupied node in a chain of $N$ nodes. The specie can be  an individual belonging to a given community or a person characterized by a certain opinion. The probabilities of choosing either of the possible specie (community birth rates) are equal to $1/2$. If $n$ nodes filled with identical species (e.g., $\downarrow \downarrow \downarrow$) get surrounded by individuals belonging to other community (e.g., $\uparrow \downarrow \downarrow \downarrow \uparrow$), the nodes inside the cluster are called {\it blocked} and they no longer interact with the rest of the system. This model can be easily extended onto a case with multiple number of different species $m \ge 2$ - then, similarly as in the two-specie case, the type of the specie is drawn from the uniform distribution $\langle 1,m \rangle$. To form an isolated cluster, a set of identical species has to be surrounded by other identical species (e.g., 23332). The analytical treatment shows that the approximated equation for the number of blocked nodes $Z$ for specific time of the simulation $t$ is
\begin{equation}\label{eq:zold}
Z \approx \frac{(m-1)t^3}{m^2 N^2}.
\end{equation}
It follows, that regardless of the number of different species introduced in the chain, the number of blocked nodes grows roughly as $t^3$. The form of Eq. (\ref{eq:zold}) gives the opportunity to directly obtain the {\it characteristic time} $t_c$ i.e., the time when the first isolated node appears 
\begin{equation}\label{eq:tcold}
t_c = \left( \frac{m^2}{m-1} \right)^{1/3} N^{2/3}.
\end{equation}
The approach is quite consistent with the numerical simulations and works also in case of other topologies such as lattices or random graphs \cite{sienkiewicz_pre}.

\section{GENERAL ANALYTICAL DESCRIPTION OF THE ASYMMETRIC CASE}\label{sec:gen_desc}
\begin{figure}
\centerline{\epsfig{file=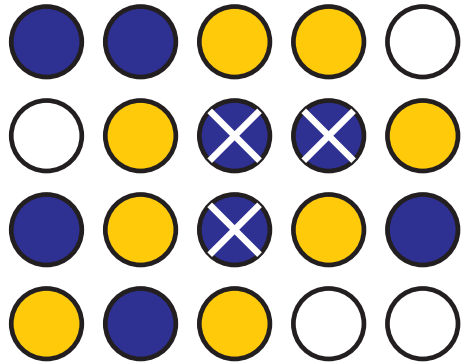,width=0.75\columnwidth}}
\caption{(Color online) An example of cluster of size 3 on square lattice ($d=2$) with $N=20$ nodes and two different species (light and dark filled circles), after $t=16$ time steps. Nodes marked with crosses are blocked.}
\label{fig:kulki}
\end{figure}

 In this paper we are investigating the asymmetric case of the model \cite{sienkiewicz_pre}. Symmetry breaking is introduced as an external bias and is received by changing birth rates of different species occurring in the  system, either by a simple imbalance in the two-specie case or by setting a specific specie probability distribution in the multi-specie one. We shall consider the model of isolation of communities on a $d$-dimensional lattice. In this kind of graphs clusters of nodes sharing the same specie will be isolated when neighbors (in the sense of von Neumann neighborhood \cite{von_neumann}) of all their elements are connected only with each other or with a different specie. An example of isolated cluster on the square lattice is presented in Fig. \ref{fig:kulki}.

\subsection{Two-specie case}
First, we consider two species ($\uparrow$ and $\downarrow$)  which we put into $d$-dimensional hypercubic lattice with a  total number of sites $N=L^d$. Occurrence probabilities  $p_\uparrow$ and $p_\downarrow$ of both species are non-equal and given by  
\begin{equation}\label{eq:peps}
	p_\uparrow = \frac{1}{2}+\epsilon, \ \ p_\downarrow = \frac{1}{2}-\epsilon,
\end{equation}
where $\epsilon \in  \langle 0,1/2 \rangle$ is the symmetry breaking parameter. These probabilities are microscopic parameters, which describe model evolution. Starting from a lattice of empty nodes, after   $t$ time steps we have the following   probabilities that a randomly picked node is occupied with the specie $\uparrow$ or $\downarrow$ 
\begin{equation}\label{eq:pr1}
	{\rm Prob}(\uparrow)=\frac{t}{N} p_\uparrow, \ \ {\rm Prob}(\downarrow)=\frac{t}{N} p_\downarrow.
\end{equation} 
\begin{figure*}
\epsfig{file=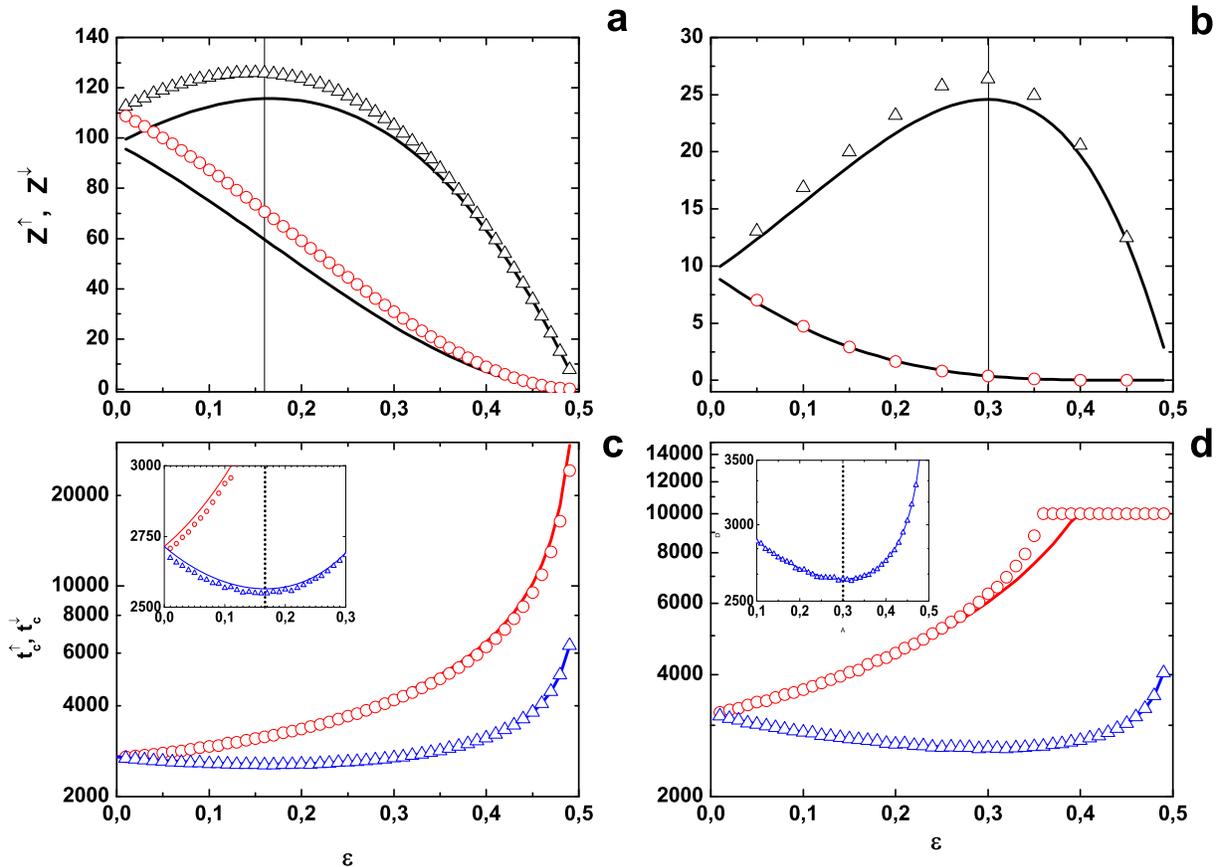,width=0.9\textwidth}
\caption{(Color online) (a-b) Number of blocked nodes $Z_{\uparrow}$ and $Z_{\downarrow}$ versus symmetry breaking parameter $\epsilon$ on a chain (a) and on a square lattice (b); circles ($Z_{\uparrow}$) and triangles ($Z_{\downarrow}$) correspond to simulation results and lines are calculated using Eqs (\ref{eq:zupdeps}-\ref{eq:zdowndeps}). (a) Simulations performed for  $t=N/4$ with $N=50000$ nodes, averaged over $100000$ runs. (b) Simulations performed for  $t=N/2$ with $N=10000$ nodes, averaged over $1000$ runs. The regions on left-hand side of the vertical line of each plot correspond to positive values of the derivative $\partial Z_{\downarrow} / \partial \epsilon$ while in those on the right-hand side $\partial Z_{\downarrow} / \partial \epsilon < 0$. The line dividing two regions is drawn for $\epsilon = \epsilon^{*}=1/6$ for the one-dimensional case and $\epsilon = \epsilon^{*} = 3/10$ for the two-dimensional  as predicted by Eq. (\ref{eq:emax}). (c-d) Log-linear plots of the characteristic times $t_c^{(\uparrow)}$ (circles) and $t_c^{(\downarrow)}$ (squares) for $d=1$ (c) and $d=2$ (d). Data points are taken from numerical simulations while solid lines come from Eqs (\ref{eq:tupd}-\ref{eq:tdownd}). (c) $N=50000$, averaged over $100000$ runs. (d) $N=10000$, averaged over $10000$ runs. The insets in both plots show in detail the range for which $t_c^{(\downarrow)}$ takes the minimum value ($\epsilon = \epsilon^{*} = 1/6$ for one-dimensional case and $\epsilon = \epsilon^{*} = 3/10$ for the two-dimensional) marked by vertical dotted lines.}
\label{fig:Z}
\end{figure*}
Now let us consider numbers of blocked  nodes  $Z^{\uparrow}$, $Z^{\downarrow}$ of both species at time $t$. These numbers  are well approximated by  numbers of individual blocked  nodes, i.e. by  numbers of blocked clusters of size one of both species. When the total density of all blocked nodes is small  ($ Z^{\uparrow}+Z^{\downarrow}\ll t$) then 
\begin{equation}\label{eq:zupd}
	Z^{\uparrow}\approx Z^{\uparrow}_1=(L-2)^d {\rm Prob}^{2d}(\downarrow) {\rm Prob}(\uparrow),
\end{equation}
\begin{equation}\label{eq:zdownd}
	Z^{\downarrow} \approx Z^{\downarrow}_1=(L-2)^d {\rm Prob}^{2d}(\uparrow) {\rm Prob}(\downarrow).
\end{equation}
Substituting the values ${\rm Prob}(\uparrow)$ and ${\rm Prob}(\downarrow)$ with Eq. (\ref{eq:pr1}) and taking into account (\ref{eq:peps}) the above relations for $L \gg 2$ may be expressed as
\begin{equation}\label{eq:zupdeps}
	Z^{\uparrow} \approx \frac{t^{2d+1}}{N^{2d}} \left( \frac{1}{4} - \epsilon^2 \right) \left( \frac{1}{2} - \epsilon \right)^{2d-1},
\end{equation}
\begin{equation}\label{eq:zdowndeps}
	Z^{\downarrow} \approx \frac{t^{2d+1}}{N^{2d}} \left( \frac{1}{4} - \epsilon^2 \right) \left( \frac{1}{2} + \epsilon \right)^{2d-1}.
\end{equation}
\begin{figure}
\centerline{\epsfig{file=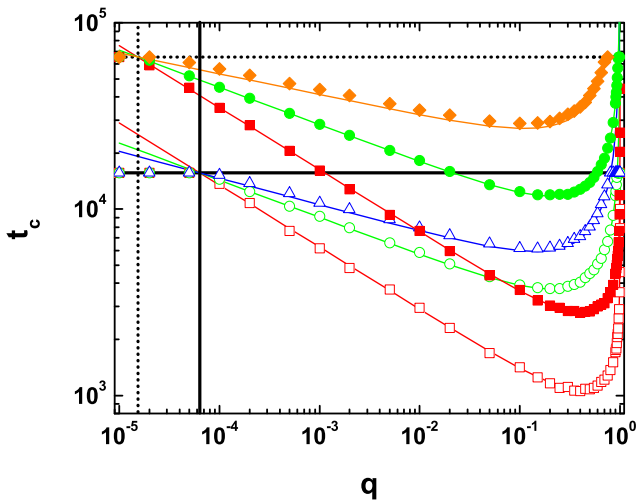,width=\columnwidth}}
\caption{(Color online) Log-log plot of the global characteristic times $t_c$ versus parameter $q$ of the geometric distribution (\ref{eq:geo}) for $N_1=15625$ (empty symbols) and $N_2=65536$ (filled symbols). Data points are taken from numerical simulations: squares are $d=1$, circles are $d=2$, triangles are $d=3$ and diamonds are $d=4$; all data points are averaged over 1000 runs. Solid lines come from Eq. (\ref{eq:tct}) and the horizontal solid line is drawn for $t_c=N_1$ whereas the horizontal dotted one for $t_c=N_2$. The vertical solid line marks $q_{min}=1/N_1$ while the vertical dotted line is drawn for $q_{min}=1/N_2$ (see description in text).}
\label{fig:tct}
\end{figure}
The above equations allow us to calculate characteristic times $t^{\uparrow}_c$ and $t^{\downarrow}_c$ when the first representative of either of both species emerges. Putting $Z^{\uparrow}=1$ (analogously for $\downarrow$) we obtain after a short algebra
\begin{equation}\label{eq:tupd}
t^{\uparrow}_c=  \left[ \frac{N^{2d}}{\left( \frac{1}{4}-\epsilon^2 \right) \left( \frac{1}{2}-\epsilon \right)^{2d-1}} \right]^{\frac{1}{2d+1}},
\end{equation} 
\begin{equation}\label{eq:tdownd}
t^{\downarrow}_c=  \left[ \frac{N^{2d}}{\left( \frac{1}{4}-\epsilon^2 \right) \left( \frac{1}{2}+\epsilon \right)^{2d-1}} \right]^{\frac{1}{2d+1}}.
\end{equation} 
\subsection{Multi-specie case}
If we consider more than two species, we can use probability distribution ${\rm Prob}(i)$ instead of the previously introduced probabilities ${\rm Prob}(\uparrow)$ and ${\rm Prob}(\downarrow)$. It requires changing microscopic probabilities $p_{\uparrow}$ and $p_{\downarrow}$ with the value of $p_i$. The connection between these two sets of variables is given by the  equation
\begin{equation}\label{eq:priapproxd}
	{\rm Prob}(i) = \frac{p_i t}{N}
\end{equation}
for $i=1,2,...$ (the case $p_i$ independent of $i$ corresponds to the symmetric problem considered in \cite{sienkiewicz_pre}). Using the same approximation as in the two-specie case we can obtain that
\begin{equation}\label{eq:zid}
	Z^{(i)}\approx Z^{(i)}_1 \approx N  \sum_{k=1,\ k\neq i}^{\infty} {\rm Prob}(k)^{2d}{\rm Prob}(i)
\end{equation}  
for $i=1,2,...$, which, taking into account (\ref{eq:priapproxd}) becomes
\begin{equation}\label{eq:ziaproxd}
	Z^{(i)} = \frac{t^{2d+1}}{N^{2d}} p_i \left( \sum_{k=1}^{\infty}p_k^{2d} -p_i^{2d} \right).
\end{equation}
Putting $Z^{(i)}=1$ one obtains a set of characteristic times for each specie $i$
\begin{equation}\label{eq:ttid}
	t_c^{(i)}= \left[ \frac{N^{2d}}{p_i \left( \sum_{k=1}^{\infty}{p_k^{2d}}- p_i^{2d} \right) } \right]^{\frac{1}{2d+1}} \ \ \ i= 1, 2, ...
\end{equation}
\section{NUMERICAL SIMULATIONS AND DISCUSSION}\label{sec:num}
Here we are presenting the comparison between the analytical approach given in the previous section and the results obtained from the numerical simulations performed for topologies of $d$-dimensional lattices. 
\subsection{Two-specie case}
While comparing Eq. (\ref{eq:zold}) for $m=2$ and Eqs. (\ref{eq:zupdeps}-\ref{eq:zdowndeps}) for specific value of $\epsilon$ one can notice that the introduction of the symmetry breaking does not have any effect on the dependence of the number of blocked nodes with regard to the time of the evolution. However, investigation of $Z^{\downarrow}$ and $Z^{\uparrow}$ as a function of the parameter $\epsilon$ for a fixed value of time $t$ brings some not so obvious results. Figure \ref{fig:Z}a-b shows the dependence of the number of blocked nodes of each specie versus the symmetry breaking parameter $\epsilon$ obtained in the numerical simulations compared with the theoretical expectations given by Eqs (\ref{eq:zupdeps}-\ref{eq:zdowndeps}). The discrepancies between simulation results and the theoretical approach can be justified by the approximations used in obtaining $Z^{\downarrow}$ and $Z^{\uparrow}$. Still, one can immediately spot the main difference between those two quantities: $Z^{\uparrow}$ is monotonic while $Z^{\downarrow}$ first increases, reaches a prominent and well defined maximum and then drops down. This observation is backed with a simple analysis of Eqs. (\ref{eq:zupdeps}-\ref{eq:zdowndeps}): in fact the derivative $\partial Z_{\uparrow} / \partial \epsilon < 0$ for the whole range $\epsilon \in \langle 0, 1/2 \rangle$ and in the case of $Z^{\downarrow}$ there is a maximum value for 
\begin{equation}\label{eq:emax}
\epsilon^{*} = \frac{1}{2} \frac{2d-1}{2d+1}.
\end{equation}
The decrease of $Z^{\uparrow}$ is rather obvious as it is the dominant specie according to Eq. (\ref{eq:peps}) - the higher is the number of its individuals introduced in the system the smaller is the probability of this specie being blocked. On the other hand as it concerns $Z^{\downarrow}$, Eq. (\ref{eq:emax}) suggests that there exists a specific value of $\epsilon$ for which the number of blocked individuals reaches the highest rate. It follows that this value depends only on the dimensionality of the system (i.e., the number of the nearest neighbors).
\begin{figure*}
\centerline{\epsfig{file=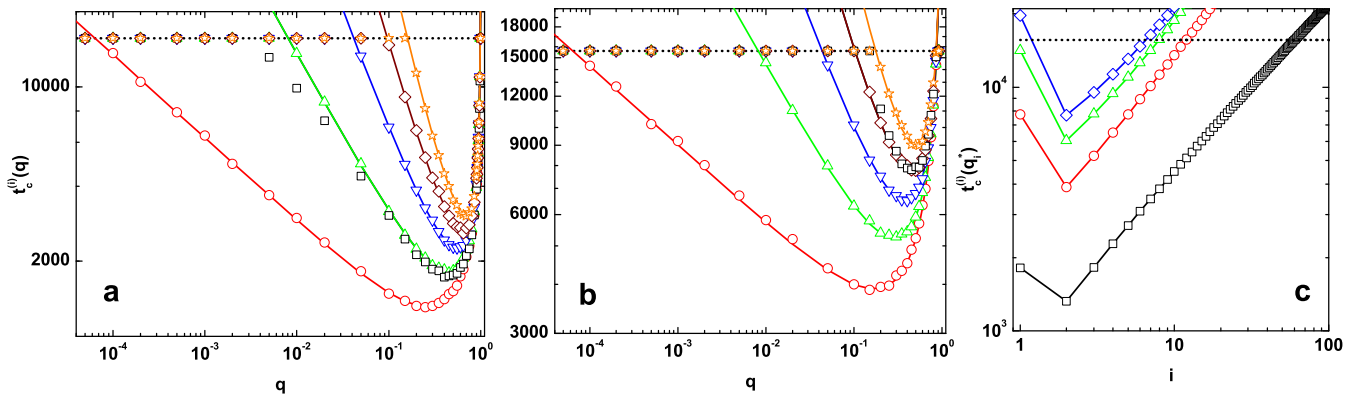,width=\textwidth}}
\caption{(Color online) (a-b) Log-log plots of characteristic times $t_c^{(i)}$ versus parameter $q$ of the geometric distribution (\ref{eq:geo}) for $d=1$ (a) and $d=2$ (b). Data points are taken from numerical simulations: squares represent specie $i=1$, circles - $i=2$, upward triangles - $i=3$, downward triangles $i=4$, diamonds - $i=4$ and stars - $i=6$; solid lines come from Eq. (\ref{eq:tcd}). Theoretical curves for $i=1$ and $i=3$ in one-dimensional case as well as $i=1$ and $i=5$ in two-dimensional overlap. In both cases $N=15625$ and the horizontal dotted line is drawn for $t_c^{(i)}=N$. Numerical data are averaged over 1000 runs. (c) Log-log plot of the characteristic time $t_c^{(i)}$ calculated from Eq. (\ref{eq:tcd1}) for $q=q^{*}_i$ against specie number $i$. Squares are $d=1$, circles are $d=2$, triangles are $d=3$ and diamonds are $d=4$; in each case $N=15625$. Solid lines are guidance to eye and the horizontal dotted line is drawn for $t_c^{(i)}=N$.}
\label{fig:tcq}
\end{figure*}
This phenomenon is even more pronounced while examining the characteristic times $t^{\uparrow}_c$ and $t^{\downarrow}_c$ versus $\epsilon$ presented in Fig. \ref{fig:Z}c-d. The first quantity exhibits a constant growth and the second one possesses a clear minimum for $\epsilon = \epsilon^{*}$ which is consistent with the maximum values for $Z^{\downarrow}$ observed in Fig. \ref{fig:Z}a-b.
\subsection{Multi-specie case}
For the asymmetric multi-specie case we used the geometric probability distribution
\begin{equation}\label{eq:geo}
p_i=q^{i-1}(1-q) \ \ \ \ \ i=1,2,...
\end{equation}
This very case of probability distribution has been chosen as an example as it quite easily yields the analytical approach. Moreover it can be transformed directly into continuous exponential distribution assuming that $q=\exp(-\alpha)$. 
First we shall discuss the issue of the global characteristic time $t_c$. It can be approximately calculated by assuming that the total number of blocked species at time $t$ is equal to $Z(t)=\sum_{i}Z^{(i)}=1$. Using Eq. (\ref{eq:geo}) the global characteristic time can be expressed as
\begin{equation}\label{eq:tct}
t_c = \left[ \frac{N^{2d}}{\frac{(1-q)^{2d}}{1-q^{2d}} - \frac{(1-q)^{2d+1}}{1-q^{2d+1}}}\right]^{\frac{1}{1+2d}}
\end{equation}
Figure \ref{fig:tct} presents the characteristic time versus parameter $q$ given for different values of $d$ and $N$. In each case the curve has a similar shape, exceeding $t_c=N$ for both small and large values of $q$ (horizontal lines in Fig. \ref{fig:tct}) with a well defined minimum between. Moreover, for a specific value of $N$ the curves, regardless of the dimensionality of the system, seem to intersect in one point. In fact, a closer analysis reveals that for $q \ll 1$ Eq. (\ref{eq:tct}) takes a form of $t_c=(N^{2d}/q)^{1/(2d+1)}$ resulting in the intersection point $q_{min}=1/N$, shown as vertical lines in Fig. \ref{fig:tct}. A heuristic explanation of this fact can be also derived in the following way: as long as the probability of drawing the second specie ($i=2$) is above $1/N$ there is a statistical chance of it appearing in the system and thus being blocked by the overwhelming first specie ($i=1$). As soon as the probability drops below that value there is only one specie and so the blocking is impossible. On the other hand, when $q$ approaches 1, the number of different species is relatively high and all the probabilities $p_i$ are close to $p_1$. After crossing a certain value $q_{max}$ the number of the individuals of the first specie is too low to make possible any blocking on the lattice. The obvious necessary condition preventing this scenario is that $N p^{2d}_1 (p_2+p_3+p_4+\dots)=1$ which leads to $N p^{2d}_1 \approx 1$ and eventually gives $q_{max} \approx 1-N^{-1/2d}$. Finally Fig. \ref{fig:tct} suggests that there is some specific value of $q$ for which the characteristic time is the lowest; assuming that $q^{2d} \ll 1$ one can estimate this value with 
\begin{equation}\label{eq:qstar}
q^{*}=\frac{1}{2d+1}.
\end{equation}
We stress here that the critical value of $q^{*}$ given by Eq. (\ref{eq:qstar}) is independent on the system size $N$ similarly as the critical value of the $\epsilon^{*}$ parameter given by Eq. (\ref{eq:emax}).

One can also focus directly on the question of characteristic times $t^{(i)}_c$ of various species. Substituting Eq. (\ref{eq:ttid}) with (\ref{eq:geo}) one gets the general formula for the characteristic time of $i$-th specie being blocked in the topology of the $d$-dimensional lattice
\begin{equation}\label{eq:tcd}
t^{(i)}_c = \left[\frac{N^{2d}(1-q^{2d})}{q^{i-1}(1-q)^{2d+1} \left(1-q^{2d(i-1)}+q^{2di} \right) }\right]^{\frac{1}{2d+1}}.
\end{equation}
The comparison of the characteristic times $t^{(i)}_c$ versus parameter $q$ obtained in numerical simulations and the theoretical expectations given by Eq. (\ref{eq:tcd}) is presented in Fig. \ref{fig:tcq}. The plots for $t^{(i)}_c$ bear close resemblance to those obtained for $t_c$. Following a similar line of thought as in the case of the global characteristic time it is possible to estimate the crucial points of these curves. First of all, Eq. (\ref{eq:tcd}) can be approximated as
\begin{equation}\label{eq:tcd1}
  \left\{
    \begin{array}{cc}
      t^{(1)}_c \approx \frac{\left( N^{2d}q^{-2d} \right)^{\frac{1}{2d+1}}}{1-q} &  \\
      t^{(i)}_c \approx \frac{\left( N^{2d}q^{1-i} \right)^{\frac{1}{2d+1}}}{1-q} & \ \ \ \ i=2,3,...,
    \end{array}
  \right.
\end{equation}
which is valid for all $q \in \langle 0,1 \rangle$. However, in order to estimate $q \ll 1$ when the above function intersects with $t^{(i)}_c=N$, one uses a further approximation i.e., $t^{(i)}_c=N^{2d}q^{(1-i)/(2d+1)}$. It effects in quite straightforward formula for the intersection point being $q_{min}=N^{1/(1-i)}$, which suggests that the intersection point should be dependent only on the number of nodes in the lattice, while its dimension $d$ is irrelevant. Figure 4 also gives clear evidence that the characteristic times for the first specie behave differently than those for $i \ge 2$, in both cases it is the specie $i=2$ which gets blocked first for smaller values of $q$. Moreover in the case of $d=1$ the results from $t^{(1)}_c$ cover with $t^{(3)}_c$ and in the case of $d=2$ with $t^{(5)}_c$. This last fact is fully comprehensible after comparing both equations in (\ref{eq:tcd1}) where, after short algebra, one gets that the characteristic time for the first and $(2d+1)$-th specie are the same. One can also notice comparing Eqs (\ref{eq:tct}) and (\ref{eq:tcd}) that it is in fact $t^{(2)}_c$ that plays the dominant role in creating the shape of the global characteristic time $t_c$ and may be used as its good approximation.  

Taking into account Eq. (\ref{eq:tcd1}) it is possible to estimate the maximal specie number that gets blocked in the environment for a fixed value of the system size $N$. The curves presented in Fig. \ref{fig:tcq}a-b indicate that for each $i$ there is a specific value $q^{*}_i$ for which the function $t^{(i)}_c$ takes its minimum. It follows, that as long as $t^{(i)}_c(q^{*}_i) < N$ the specie will be blocked, at least for $q=q^{*}_i$. Closer analysis of Eq. (\ref{eq:tcd1}) leads to 
\begin{equation}\label{eq:qstari}
  \left\{
    \begin{array}{cc}
      q^{*}_1 = \frac{2d}{4d+1} &  \\
      q^{*}_i = \frac{i-1}{i+2d} & \ \ \ \ i=2,3,...,
    \end{array}
  \right.
\end{equation}
After substituting Eq. (\ref{eq:tcd1}) with the above values of $q^{*}_i$, one gets the value of characteristic time in the minimum. A corresponding plot for different values of dimension $d$ is shown in Fig. \ref{fig:tcq}c. It gives immediately the idea of the fast restriction of the blocked specie number with system's dimension: while for $d=1$ the value of $i$ can be as large as 70, in the case of $d=4$ it substantially drops down to 6. Furthermore, Fig. \ref{fig:tcq}c underlines again the specific role of the first specie showed in the previous paragraph.
\section{CONCLUSIONS}\label{sec:conc}
In this paper we have extended the simple model of community isolation onto the case with the symmetry breaking. Our calculation and numerical simulations show that even a simple way of introducing the external bias between species can lead to interesting and non-trivial results. We have found that both in the two-specie case where a parameter $\epsilon$ governs the symmetry breaking and in the multi-specie case where the numbers of each entity are given by the geometric distribution there exists some specific and well defined value of the control parameter giving a minimum of the characteristic time $t_c$. In general the value of this parameter is dependent only on the dimensionality of the lattice. We have also shown that the requirement of a non-blocked system in the multi-specie case leads to two, somehow opposite conditions: either there has to be a strictly dominant specie, outnumbering the others or the diversity should be very large. In the end we have also given evidence that the number of blocked species of different kinds decreases with the dimension of the system. The presented results can be easily further generalized to the cases of other topologies (e.g. complex networks) and other kinds of biases.
\begin{acknowledgments}
The authors acknowledge support from European COST Action MP0801 Physics of Competition and Conflicts and from Polish Ministry of Science Grant 578/N-COST/2009/0.
\end{acknowledgments}

\end{document}